\documentclass[12pt]{iopart} 
\usepackage{iopams}
\usepackage[applemac]{inputenc}
\def\eps{\varepsilon}
\newcommand{\eqref}[1]{(\ref{#1})}

\begin{document}
\title[Statistics of non-interacting bosons and fermions]{Statistics of non-interacting bosons and fermions in 
micro-canonical, canonical and grand-canonical ensembles: A survey}
\author{ Fabrice Philippe \dag\ddag ,
Jacques Arnaud \P\ and 
Laurent Chusseau $\parallel$}
\address{\dag\ Université Paul Valéry, 
Route de
Mende, F-34199 Montpellier Cedex 5}
\address{\ddag\ LIRMM, 161 rue Ada, F-34392 
Montpellier}
\address{\P\ Mas Liron, F-30440 Saint Martial, 
France}
\address{$\parallel$\ Centre
d'Electronique et de Micro-optoélectronique de Montpellier, Unité 
Mixte de
Recherche n°5507 au CNRS, Université Montpellier II, F-34095 
Montpellier,
France}
\ead{fabrice.philippe@univ-montp3.fr}
\begin{abstract}
The statistical properties of
non-interacting bosons and fermions confined in trapping potentials
are most easily obtained when the system may exchange energy and
particles with a large reservoir (grand-canonical ensemble).  There
are circumstances, however, where the system under consideration may
be considered as being isolated (micro-canonical ensemble).  This
paper first reviews results relating to micro-canonical ensembles. 
Some of them were obtained a long time ago, particularly by Khinchin
in 1950.  Others were obtained only recently, often motivated by
experimental results relating to atomic confinement. A number of formulas are
reported for the first time in the present paper. Formulas
applicable to the case where the system may exchange energy but not
particles with a reservoir (canonical ensemble) are derived from the
micro-canonical ensemble expressions. The differences
between the three ensembles tend to vanish in the so-called
Thermodynamics limit, that is, when the number of particles and the volume go to
infinity while the particle number density remains constant.  But we 
are mostly interested in systems of moderate size, often referred to 
as being \emph{mesoscopic}, where the grand-canonical formalism is not 
applicable. The mathematical results rest primarily on the enumeration of partitions 
of numbers. 
\end{abstract}
\submitto{\JPA}

\section{Introduction}

Let us first recall that in Quantum Statistical Mechanics isolated systems
should be treated in the \emph{microcanonical} formalism, systems that may
exchange energy with a heat bath in the \emph{canonical}
formalism and systems that may exchange both energy and particles
with a bath in the \emph{grand-canonical} formalism.  In
the latter case simple formulas are easily obtained.  For large
systems the three ensembles are generally equivalent so that the
grand-canonical formalism suffices for most applications.  However,
there is presently much interest in systems consisting of only a few
particles, bosons or fermions. Recent experimental observations of atomic
Bose-Einstein condensation motivated much theoretical work
relating to \emph{mesoscopic} devices.  The fluctuation of the
number of bosons occupying the ground level of harmonic traps was
particularly considered.  In such cases the grand-canonical ensemble
is inappropriate and one must consider instead the less-easily handled 
micro-canonical and canonical formalisms. 

Many results about canonical and microcanonical ensembles are known 
but they are scattered in the literature and often require intricate
mathematical methods.  The present paper provides an essentially
exhaustive and mathematically simple derivation of the statistics of
non-interacting particles in the three ensembles.  Fermions
and bosons are treated on the same footing but particles such as
photons and phonons that do not carry generalized charges are
considered in a separate section.  In our approach
\emph{system} energy levels and their degeneracies are considered.  A short study of
compound systems helps clarify how degeneracy occurs.  The
mathematical method rests on exact results relating to the
microcanonical statistics, canonical and grand-canonical expressions
being subsequently derived through averaging.  This line of attack is
often evoked as being the most natural one but it has apparently not
been pursued in detail before.  Microcanonical statistics was 
treated by Khinchin in 1950 \cite{kn:k51} but other statistics and 
degeneracies were not considered at that time.
The microcanonical theory rests on the assumptions that the particles are undistinguishable and
interact only weakly so that one-particle energies add up. The 
 number of
microcanonical configurations is assumed to be finite.

The purpose of this paper is not purely educational.  Although the
mathematical formalism requires only elementary combinatorics and the
handling of generating series, the method naturally gives rise to
several new results. This is the case in particular
for formulas that facilitate estimations of certain quantities relevant to the
canonical and grand-canonical ensembles.  Complete results are given
in the important case of evenly-spaced non-degenerate levels,
one-dimensional harmonic potentials being a basic example. 
Recursive relations for occupancies are obtained that generalize
results recently reported by Sch\"{o}nhammer \cite{kn:s00}. It has
been noticed long ago that microstates are in that case
described by the partition of integers \cite{kn:vlu37, kn:h38}. It
is also known that hard-core interacting bosons in harmonic potentials
may be treated as non-interacting fermions \cite{kn:g65}. 
Furthermore, the micro-canonical statistics of a band of states is
particularly relevant to spin systems \cite{kn:pa01} and semiconductor
devices.

The paper is essentially self-contained. Various formulas about
occupancies, statistical weights, and partition functions for bosons and
fermions are obtained in Sections~2 to 4, beginning with the microcanonical
and ending up with the grand-canonical ensemble.  Chargeless particles are 
treated in Section 5, and  Section 6 addresses issues relating to compound systems and 
degeneracy. Section 7 deals with the case of evenly-spaced levels.  Relations between fermionic and bosonic statistics are given in  Appendix A, links between von Neumann-Shannon entropy and the classical formulations of entropy in Thermodynamics in Appendix B. 

\section{Microcanonical statistics}\label{sec2}

Consider an isolated system comprising $N$ undistinguishable particles.  The energy $\eps_s$ of a particle in (quantum)
state $s$ is assumed to be non-negative without loss of generality. 
In what follows $\mathbb{S}$ denotes the set of states and $\mathbb{E}$
the set of energy levels.  The \emph{degeneracy} $g_{\eps}$ of an
energy level $\eps$ is the number of states corresponding to that level.  For
example the energy of a state $s\equiv(k_1,\ldots, k_d)$, where $k_1,\ldots, k_d$ are non-negative integers, of a $d$-dimensional isotropic harmonic
oscillator is given by the formula $\eps=\hbar\omega(k+d/2)$
where $\hbar\omega$ is a constant and $k=k_1+\cdots+k_d$. The degeneracy 
of the energy level $\eps$ is equal to the number
of ways of obtaining $k$ by summing up $d$ nonnegative integers,
namely $g_{\eps}=(k+d-1)!/(k!(d-1)!)$.

A system is characterized by $N$ and the total energy $U$ it contains. Because of our assumption that the 
one-particle energies add up,  $U$ is the sum of $N$ 
$\eps$-values ($\eps\in\mathbb{E} $), some of them possibly occuring more than once.
In the sequel $\mathbb{U}$ denotes the set 
of \emph{all} finite sums of elements of $\mathbb{E}$, with the 
convention that $\mathbb{U}$ 
contains $0$. Given $N\in\mathbb{N}$ and $U\in\mathbb{U}$ several 
configurations may occur when $U$ is a sum of $N$ values in $\mathbb{E}$. 
 Letting $\nu_{s}$ denote the number of particles in state $s$, a
$(N,U)$-microstate is a family 
$\nu=(\nu_{s})_{s\in\mathbb{S}}$ of non-negative integers (in other words, $\nu$ is a mapping from $\mathbb{S}$ to $\mathbb{N}$) fullfilling the conditions 
\begin{equation}\label{defm}
\sum_{s\in\mathbb{S}}\nu_{s}=N,\quad\mathrm{and}\quad
\sum_{s\in\mathbb{S}}\nu_{s}\eps_s=U.
\end{equation}
 
\begin{figure}[h]
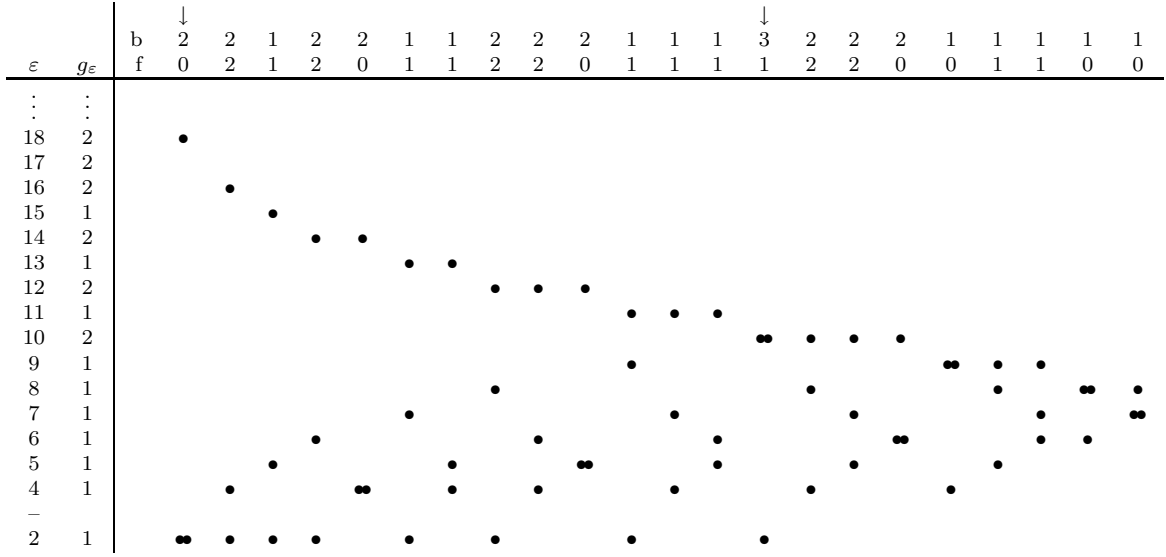
 
	\begin{center}	
	\begin{scriptsize}
	\bigskip
	\begin{tabular}{cc|rcccccccccccccccccccccc}
&&&$\downarrow$&&&&&&&&&&&&&$\downarrow$\\
&&b&2&2&1&2&2&1&1&2&2&2&1&1&1&3&2&2&2&1&1&1&1&1\\
$\eps$&$g_\eps$&f&0&2&1&2&0&1&1&2&2&0&1&1&1&1&2&2&0&0&1&1&0&0\\
\hline\vdots&\vdots&\\
18&2&&$\bullet$\\
17&2&&\\
16&2&&&$\bullet$\\
15&1&&&&$\bullet$\\
14&2&&&&&$\bullet$&$\bullet$\\
13&1&&&&&&&$\bullet$&$\bullet$\\
12&2&&&&&&&&&$\bullet$&$\bullet$&$\bullet$\\
11&1&&&&&&&&&&&&$\bullet$&$\bullet$&$\bullet$\\
10&2&&&&&&&&&&&&&&&$\bullet\!\bullet$&$\bullet$&$\bullet$&$\bullet$\\
9&1&&&&&&&&&&&&$\bullet$&&&&&&&$\bullet\!\bullet$&$\bullet$&$\bullet$\\
8&1&&&&&&&&&$\bullet$&&&&&&& 
$\bullet$&&&&$\bullet$&&$\bullet\!\bullet$&$\bullet$\\
7&1&&&&&&&$\bullet$&&&&&&$\bullet$&&&&$\bullet$&&&&$\bullet$&&$\bullet\!\bullet$\\
6&1&&&&&$\bullet$&&&&&$\bullet$&&&&$\bullet$&&&&$\bullet\!\bullet$&&&$\bullet$&$\bullet$\\
5&1&&&&$\bullet$&&&&$\bullet$&&&$\bullet\!\bullet$&&&$\bullet$&&&$\bullet$&&&$\bullet$\\
4&1&&&$\bullet$&&&$\bullet\!\bullet$&&$\bullet$&&$\bullet$&&&$\bullet$&&&$\bullet$&&&$\bullet$\\
--&&\\
2&1&&$\bullet\!\bullet$&$\bullet$&$\bullet$&$\bullet$&&$\bullet$&&$\bullet$&&&$\bullet$&&&$\bullet$\\ 

\end{tabular} \caption{This figure represents the ($N=3,~U=22$)-microstates: 
The energy levels $\eps$ and 
their degeneracy $g_\eps$ are shown on the left. At the top of the 
figure the 
number of microstates corresponding to each partition is given for bosons (b) and fermions (f).} \label{fig1}
	\end{scriptsize}
	\end{center}
\end{figure}

As an example consider a system of particles whose energy levels are 
$\eps =2k-1+l^2$, with $k,l$ positive integers (this is the 
case for charged single-spin particles in uniform magnetic fields 
under some conditions).  Figure \ref{fig1} shows the 22 
partitions of $U=22$ among $N=3$ particles.  Because of the
degeneracies some of the partitions shown are associated with \emph{several}
microstates. 
The number $\nu_{s}$ of bosons in some state $s$ is unconstrained while the 
numbers of fermions in a state may only be 0 or 1.  Accordingly, the first 
marked ($\downarrow$) partition corresponds to two bosonic microstates since
the $\eps$=18 level is twice degenerate and to zero fermionic
microstate since the $\eps$=2 level is not degenerate. The second marked 
partition in Fig. \ref{fig1} may correspond to a fermionic microstate
because two \emph{distinct} states have energy $\eps$=10, namely
$s_{1,3}$ (i.e., $k$=1, $l$=3) and $s_{5,1}$ (i.e., $k$=5, $l$=1), or 
to three bosonic microstates where the two-particles states for 
$\eps$=10 are respectively: $(s_{1,3},s_{5,1})$,
$(s_{1,3},s_{1,3})$ and  $(s_{5,1},s_{5,1})$. 

The system \emph{statistical weight} $W(N,U)$, i.e., the number of
distinct $(N,U)$-microstates, is assumed to be finite. This implies 
finite degeneracies.  Indeed, $W(1,U)$ is zero if $U$ is not a one-particle energy level, otherwise it is the degeneracy at this level.  Notice that since $0\in\mathbb{U}$ by convention $W(0,0)=1$.  Summing up the numbers on top of Fig. \ref{fig1} we see that $W(3,22)$ is equal to 34 for bosons and 21 for fermions. The purpose of formulas given below is to provide such numbers directly.

 The \emph{microcanonical
occupancy} $N_\eps(N,U)$ of energy level $\eps$ is defined as 
the average number of particles at this energy level, the $W(N,U)$ microstates being equally likely. (If $W(N,U)= 0$ we have
$N_\eps(N,U)=0$). In Fig. \ref{fig1} the occupancy of level $\eps=10$ is $12/34$ for bosons and $6/21$ for fermions. Letting $\Omega(N,U)$ denote the set of all 
$(N,U)$-microstates the occupancy may be written 
\begin{equation}\label{defo}
N_\eps(N,U)=\frac{1}{W(N,U)}\sum_{s,\,\eps_s=\eps}
\sum_{\nu\in\Omega(N,U)}\nu_{s}, 
\end{equation}
the first summation index $s$ running through the states of energy $\eps$. 
Conditions \eqref{defm} and definition \eqref{defo} result in 
\begin{equation} \label{NUtot}
\sum_{\eps\in\mathbb{E}}N_\eps(N,U)=N,\quad\mathrm{and}\quad   
\sum_{\eps\in\mathbb{E}}N_\eps(N,U)\eps=U. 
\end{equation}
For example, the first relation in \eqref{NUtot} derives from  
\[
\fl W(N,U)\sum_{\eps\in\mathbb{E}}N_\eps(N,U)=
\sum_{\eps\in\mathbb{E}}\sum_{s,\,\eps_s=\eps}\sum_{\nu\in\Omega(N,U)}\nu_{s}
=\sum_{s\in\mathbb{S}}\sum_{\nu\in\Omega(N,U)}\nu_{s}
=\sum_{\nu\in\Omega(N,U)}\sum_{s\in\mathbb{S}}\nu_{s}.
\]
Notice that at most $NW(N,U)$ summands contribute to each of the above sums. 
In the sequel the symbol $+$ refers to bosons and the symbol $-$ to fermions.
The symbol $\pm$ should be replaced by $+$ for bosons and by $-$ for fermions. 

The present article rests mainly on the following recursion relation, to 
be proven later in this section
\begin{equation}
N_\eps^\pm(N,U) =
\frac{W^\pm(N-1,U-\eps)}{W^\pm(N,U)}\left (g_\eps\pm 
N_\eps^\pm(N-1,U-\eps)\right ). 
\label{mcor} 
\end{equation}
Since $W(N-n,U-n\eps)$ vanishes for $n>N$, iterating $N$ times the 
above formula yields
\begin{equation}
N_\eps^\pm(N,U) =
\frac{g_\eps}{W^\pm(N,U)}\sum_{n=1}^N(\pm 1)^{n-1}W^\pm(N-n,U-n\eps). 
\label{mco} 
\end{equation}
Combining \eqref{NUtot} and \eqref{mco} provides recursive formulas for the statistical weight
\begin{eqnarray}
W^\pm(N,U)&=\frac{1}{N}\sum_{n=1}^N\sum_{\eps\in\mathbb{E}}
(\pm 1)^{n-1}g_\eps W^\pm(N-n,U-n\eps)
\label{recW}\\
&=\frac{1}{U}\sum_{n=1}^N\sum_{\eps\in\mathbb{E}}
(\pm 1)^{n-1}\eps g_\eps W^\pm(N-n,U-n\eps). \label{recWU}
\end{eqnarray}
For example, the first one obtains as follows: 
\[\fl 
NW^\pm(N,U)=
\sum_{\eps\in\mathbb{E}}N_\eps^\pm(N,U)W^\pm(N,U)=
\sum_{\eps\in\mathbb{E}}g_\eps\sum_{n=1}^N
(\pm 1)^{n-1}W^\pm(N-n,U-n\eps).
\]
Notice that the sums above involve only the values of $\eps$ for which
$N_\eps(N,U)\neq 0$, that is, at most $NW(N,U)$ $\eps$-values. Formula 
\eqref{recW}  appears in \cite{kn:ww97,kn:cmz99} in a different form 
(summation is done on one-particle states instead of system 
energy levels) for bosons, and in \cite{kn:acp00} for fermions 
without degeneracy. Although less interesting than \eqref{recW} for a direct computation of $W(N,U)$, formula \eqref{recWU} will prove useful later.  

To establish \eqref{mcor} consider a 
state $s$.  For bosons \eqref{defm} shows that the relation
$\nu_{s}=\nu'_{s}-1$ defines a one-to-one correspondence between the 
set of $(N-1,U-\eps)$-microstates  $\nu$ and the set  of 
$(N,U)$-microstates $\nu'$ with $\nu'_{s}>0$. Thus 
\begin{equation}\fl 
\sum_{\nu'\in\Omega^+(N,U)}\nu'_{s}=
\sum_{\nu\in\Omega^+(N-1,U-\eps)}(1+\nu_{s})=
W^+(N-1,U-\eps)+\sum_{\nu\in\Omega^+(N-1,U-\eps)}\nu_{s}. 
\label{e1} 
\end{equation}
The above result amounts to removing one particle in 
state $s$ from each $(N,U)$-microstate that exhibits such a 
particle and counting the remaining particles in state $s$. 
For fermions the number of $(N,U)$-microstates with one
particle in state $s$ equals the number of $(N-1,U-\eps)$-microstates
with no particle in state $s$, that is, all of them except those 
having one particle in that state.  Thus
\begin{equation}
\sum_{\nu'\in\Omega^-(N,U)}\nu'_{s}=
W^-(N-1,U-\eps)-\sum_{\nu\in\Omega^-(N-1,U-\eps)}\nu_{s}. 
\label{e2} 
\end{equation}
By \eqref{defo}, summing extremal members in \eqref{e1} and 
\eqref{e2} for all
states $s$ of energy $\eps$ the result in \eqref{mcor} is obtained. 

Formulas \eqref{mcor} and \eqref{mco} have apparently not been
published before in their general form.  Related expressions are given
by Khinchin \cite{kn:k51} under the assumption of evenly-spaced
non-degenerate energy levels.  Khinchin accounts for degeneracy by
repeating energy levels.  This amounts to consider quantum states
instead of energy levels \cite{kn:ww97}.  For bosons Eq.  \eqref{mco}
has appeared in a different form in \cite{kn:ww97}.  For the case of
non-degenerate fermionic systems the formulas (\ref{mcor},\ref{mco}) 
were given in \cite{kn:acp00}.

It is thus clear that microcanonical statistics rely on the sole 
knowledge of statistical weights, the computation of which  
requiring only the energy-levels values and their 
degeneracy. 

\section{Canonical statistics}

When the system may exchange energy with 
a medium of arbitrarily large size
at temperature $T$ (heat bath), the probability that a particular 
$(N,u)$-microstate occurs at equilibrium is proportional to $q^u$, where
\[
q\equiv\exp(-\frac{1}{k_{\mathrm{B}}T}),
\]
according to Boltzmann and Gibbs.
The \emph{partition function} 
\begin{equation}\label{Zdef}
Z(N,q)\equiv\sum_{u\in\mathbb{U}}W(N,u)q^u
\end{equation}
normalizes the probability.  Notice that, even though all sums are 
finite in the
previous section, the summation index $u$ takes on infinitely 
many contributing values. Summability requires that the energy levels $\eps$ 
form a finite or unbounded countable set. If 
the levels are discrete this means that they must form a
finite or unbounded increasing sequence. 
Convergence of the partition function occurs if, and only if, $Z(1,q)$ converges. Indeed, $0\leq Z^-(N,q)\leq Z^+(N,q)\leq Z(1,q)^N$. We assume in the sequel that $Z(1,q)$ converges. 

The system is characterized by the number $N$ of particles and the 
temperature $T$, or $N$ and $q$ equivalently. The average energy 
$U(N,q)$ follows from 
\begin{equation}\label{avU}
U(N,q)=\frac{1}{Z(N,q)}\sum_{u\in\mathbb{U}}uW(N,u)q^u
=q\frac{d}{dq}\ln Z(N,q). 
\end{equation}
The system may be characterized by $N$ and $U$ since 
$U(N,q)$ is an increasing function of $q$. Indeed, the energy variance, a positive quantity, is given by 
\begin{eqnarray*}
\mathrm{Var}(U)&=
\frac{1}{Z(N,q)}\sum_{u\in\mathbb{U}}u(u-1)W(N,u)q^u+U(N,q)-U(N,q)^2\\
&=\frac{1}{Z(N,q)}\frac{d^2}{dq^2}Z(N,q)+U(N,q)-U(N,q)^2
\quad =q\frac{d}{dq}U(N,q).
\end{eqnarray*}
Energy variance is therefore linked to heat capacity $c(N,T)\equiv 
dU(N,q)/dT$ by $\mathrm{Var}(U)=k_{\mathrm{B}}T^2c(N,T)$. It is 
worthwhile to point out that two systems in the 
canonical ensemble have the same heat capacity if, and only if, their 
partition functions coincide except for a multiplicative factor of 
the form $aq^b$, where $a$ and $b$ are real numbers that do not 
depend on $T$.  This result follows easily from \eqref{avU}. 

The \emph{canonical occupancy} $N_\eps(N,q)$ at level $\eps$ is the 
average
number of particles at this level.  It may therefore be calculated by 
averaging
the microcanonical occupancies at level $\eps$ with $W(N,u)q^u$ as a 
weight. 
Accordingly, relations \eqref{NUtot} result in 
\begin{equation}\label{cdef}
\sum_{\eps\in\mathbb{E}}N_\eps(N,q)=N,\quad\mathrm{and}\quad 
\sum_{\eps\in\mathbb{E}}N_\eps(N,q)\eps=U(N,q). 	
\end{equation} 
Since $W(N,u)$ vanishes for $u<0$, it is not difficult to see 
\footnote{Index shifts in the calculation, such as
$\sum_{u\in\mathbb{U}}W(N-1,u-\eps)q^{u}=
q^\eps\sum_{u\in\mathbb{U}}W(N-1,u)q^{u}$, 
are justified by the following remark.  Let $u'\in\mathbb{U}$.  If 
there exists $N\in\mathbb{N}^*$ such that $W(N,u-u')\neq
0$, then $u\in\mathbb{U}$ if, and only if, $u-u'\in\mathbb{U}$. }  
with the help of formula \eqref{mcor} that
\begin{eqnarray}
N_\eps^\pm(N,q)&=q^\eps \frac{Z^\pm(N-1,q)}{Z^\pm(N,q)}
  \left (g_\eps\pm N_\eps^\pm(N-1,q)\right ), 
  \label{cor}\\
&=\frac{g_\eps}{Z^\pm(N,q)}
  \sum_{n=1}^N(\pm 1)^{n-1}q^{n\eps}Z^\pm(N-n,q).
  \label{co}	
\end{eqnarray}
Equation \eqref{co} is derived by iterating \eqref{cor} or
averaging \eqref{mco}.  A relation similar to \eqref{cor} is
given in \cite{kn:s89} for the case of non-degenerate energy levels. 
But the assumption made in that paper that 
$N_\eps^\pm(N,q)$ and $N_\eps^\pm(N-1,q)$ are identical is valid for large $N$-values only. The relations \eqref{co} and \eqref{mco} were given in \cite{kn:bhmh99}.

In general, a closed form expression of $Z(N,q)$ has not been 
established, but this quantity 
can be evaluated recursively. Using $W(0,0)=1$ and $W(0,u)=0$ 
if $u>0$, we obtain from \eqref{recW}
\begin{equation}\label{Z0Z1}
 Z(0,q)=1,\quad\mathrm{and}\quad 
Z(1,q)=\sum_{\eps\in\mathbb{E}}g_\eps q^{\eps}. 
\end{equation}
Summing up \eqref{co} for $\eps$ running through
$\mathbb{E}$ and making use of the first formula in \eqref{cdef}, 
we obtain for $N\geq 1$ 
\begin{equation}\label{recZ}	
 Z^\pm(N,q)=\frac{1}{N}
 \sum_{n=1}^N(\pm 1)^{n-1}Z^\pm(N-n,q)Z(1,q^n).
\end{equation}
This formula may also be derived directly from \eqref{recW}. 
Alternative proofs are given in \cite{kn:l61, kn:bf93, kn:ss98, kn:b01}. 
Making use of \eqref{recWU} instead provides a similar formula for the
computation of $Z'(N,q)\equiv\frac{d}{dq}Z(N,q)$.  Comparing this
formula with the derivative of \eqref{recZ} results in an other
expression.  These new formulas are respectively
\begin{eqnarray}	
 Z'^{\pm}(N,q)&=\sum_{n=1}^N(\pm q)^{n-1}Z^\pm(N-n,q)Z'(1,q^n)\\
 &=\frac{1}{N-1}\sum_{n=1}^{N-1}(\pm 1)^{n-1}Z'^\pm(N-n,q)Z(1,q^n). 
\label{recZ'}
\end{eqnarray} 
The above relations \eqref{recZ} and \eqref{recZ'} allow, in turn, the 
numerical computation of $U(N,q)$ with the help of \eqref{avU}. 
 
Note incidentally that $Z(1,q)$ depends only on the energy levels 
and their degeneracy, not on the nature of the particles.  Moreover, formulas \eqref{recZ} and \eqref{co} show that the main difficulty in the computation of $Z(N,q)$ and $N_\eps(N,q)$ rests on the numerical evaluation of 
$Z(1,q^n)$, that is, of the one-particle partition function at temperature $T/n$, for $n$ going from 1 to $N$. There is a formula giving $Z(N,q)$ in terms of $Z(1,q)$ only: 
\begin{equation}\label{iteZ}
  Z^\pm(N,q)=\sum_{(\nu_k)_{k\geq 1}\in\mathbf{P}(N)}\prod_{k\geq 1}
  \frac{(\pm 1)^{k-1}Z(1,q^k)^{\nu_k}}{\nu_k!\, k^{\nu_k}},
\end{equation} 
where $\mathbf{P}(N)$ stands for the set of unrestricted 
partitions of $N$, corresponding to the non-degenerate bosonic 
$(N,U)$-microstates with $\mathbb{E}=\mathbb{N}$ and $U=N$; see, 
e.g., \cite{kn:a76}. This formula, which may be established 
by induction on $N$ (the proof is tedious but obvious), is a 
special case of more general relations that 
are not relevant to this paper.  Related formulas are given in 
\cite{kn:ss98,kn:ss99} and linked to symmetric polynomials in \cite{kn:ss01}. 

\section{Grand-canonical statistics}
If the system may exchange energy and particles with a bath at
temperature $T$ and fugacity $z$, the probability that 
a $(n,u)$-microstate occurs is proportional to 
$z^nq^u$. The grand-partition function 
\[
\mathcal{Z}(z,q)=\sum_{n\geq 0}\sum_{u\in\mathbb{U}}W(n,u)z^nq^u 
\] 
that normalizes the probability reads 
   \begin{equation} \label{sgqtW}
   \mathcal{Z}^+(z,q)=\prod_{\eps\in\mathbb{E}}\frac{1}{(1-zq^\eps)^{g_\eps}},
  \qquad
   \mathcal{Z}^-(z,q)=\prod_{\eps\in\mathbb{E}}(1+zq^\eps)^{g_\eps}. 
   \end{equation}
These formulas can be recovered by averaging both sides of
\eqref{recZ} with $z^N$ as a weight and using \eqref{Z0Z1} so that 
\begin{eqnarray*}
\sum_{n\geq 0}nZ^\pm(n,q)z^n & =\sum_{n\geq 0}
\sum_{i=1}^n(\pm 1)^{i+1}Z^\pm(n-i,q)\sum_{\eps\in\mathbb{E}}g_\eps 
q^{i\eps}\\ 
& = -\sum_{\eps\in\mathbb{E}}g_\eps\sum_{n\geq 0}z^n
\sum_{i=1}^n(\pm q^\eps)^iZ^\pm(n-i,q)\\ 
& = -\sum_{\eps\in\mathbb{E}}g_\eps\Bigl(\sum_{n\geq 1}(\pm 
zq^\eps)^n\Bigr)
\Bigl(\sum_{n\geq 0}Z^\pm(n,q)z^n\Bigr)\\ 
&=z\sum_{n\geq 0}Z^\pm(n,q)z^n\sum_{\eps\in\mathbb{E}}g_\eps
\frac{\mp q^\eps}{1\mp zq^\eps}. 
\end{eqnarray*}
Dividing both sides by $z\sum_{n\geq 0}Z^\pm(n,q)z^n$ we 
recognize 
that the two sides of each equality shown in \eqref{sgqtW} have the same 
logarithmic derivative with respect to $z$, so that they are proportional to one another. It follows that they are identical since they have the same constant term 1. 

The \emph{grand-canonical} occupancy at level $\eps$ is calculated by
averaging the microcanonical occupancy $N_\eps(n,u)$ with
$z^nq^uW(n,u)$ as a weight, that is, by averaging the canonical
occupancy $N_\eps(n,q)$ with $Z(n,q)z^n$ as a weight.  Using
\eqref{cor} and introducing the chemical potential $\mu\equiv k_{\mathrm{B}}T\ln
z$, we obtain the celebrated Bose-Einstein and Fermi-Dirac
distributions
\begin{equation}
N_\eps^+(z,q)=\frac{g_\eps}{q^{\mu -\eps}-1}, \qquad 
N_\eps^-(z,q)=\frac{g_\eps}{q^{\mu -\eps}+1}.
\label{BE-FD} 
\end{equation}

It follows from 
$\ln\mathcal{Z}^\pm(z,q)=\mp\sum_{\eps\in\mathbb{E}}g_\eps\ln(1\mp 
zq^\eps)$ that the average number of particles and
average energy are  respectively 
\begin{equation}
\sum_{\eps\in\mathbb{E}}N_\eps(z,q)=z\frac{d}{dz}\ln\mathcal{Z}(z,q),
\qquad
\sum_{\eps\in\mathbb{E}}\eps 
N_\eps(z,q)=q\frac{d}{dq}\ln\mathcal{Z}(z,q).
\end{equation}
The expansion of $\ln(1\mp zq^\eps)$ in power series of $zq^\eps$ provides with 
an expression of $\ln\mathcal{Z}(z,q)$ in terms of the 
one-particle canonical partition function, namely 
\begin{equation}
\ln\mathcal{Z}^\pm(z,q)=\sum_{n\geq 1}(\pm 
1)^{n-1}Z(1,q^n)\frac{z^n}{n}. 
\end{equation}
Accordingly, the average number of particles and energy may be computed as 
\begin{eqnarray}
N(z,q)&=\sum_{n\geq 1}(\pm 1)^{n-1}Z(1,q^n)z^n,
\\
U(z,q)&=\sum_{n\geq 1}(\pm 1)^{n-1}Z'(1,q^n)(zq)^{n}. 
\end{eqnarray}

\section{Photons and phonons.} 
Particles that do not bear generalized charge such as photons or 
phonons can be created 
or annihilated without violating conservation rules. 
Accordingly, the number $N$ of such particles is not conserved in the 
micro-canonical and 
canonical ensembles. A 
$U$-microstate is now a family
$\nu=(\nu_{s})_{s\in\mathbb{S}}$ of non-negative integers
satisfying the sole condition
\begin{equation}\label{defmc}
\sum_{s\in\mathbb{S}}\nu_{s}\eps_s=U.  
\end{equation}
 The microcanonical statistical weight $W(U)$ is obtained by 
summing up $W(n,U)$ over all $n$ values. Let
$\eps_0=\min\mathbb{E}$ be the lowest one-particle energy level.  If 
$\eps_0 >0$
then $n\eps_0$ exceeds $U$ at sufficiently large values of $n$ so that $W(U)$ is 
finite.  This arguments also applies to fermionic systems with 
$\eps_0 =0$ because the number of fermions that occupy this level cannot exceed its degeneracy. But
if $\eps_0$ were equal to zero in bosonic systems the addition of a particle at 
level $\eps_0$ would entail that $W(n,U)\leq W(n+1,U)$ and $W(U)$ would be equal to zero or infinity. We therefore assume that the ground-state energy of chargeless bosons is non-zero. No generality is lost since such particles cannot be detected at zero energy. 

The microcanonical occupancy of the energy level $\eps$ is the 
average number of particles at level $\eps$. Arguments similar to the ones given in Section \ref{sec2} show that \eqref{NUtot}-\eqref{mco} may be written as 
\[
\sum_{\eps\in\mathbb{E}}N_\eps(U)\eps =U, 
\]
\begin{eqnarray}
N_\eps^\pm(U)&=\frac{W^\pm(U-\eps)}{W^\pm(U)}\bigl(g_\eps\pm 
N_\eps^\pm(U-\eps)\bigr)\\
&=\frac{g_\eps }{W^\pm(U)}\sum_{n=1}^{\lfloor U/\eps\rfloor}(\pm 
1)^{n-1}W^\pm(U-n\eps). 
\end{eqnarray}
Equation \eqref{recWU} becomes 
\begin{equation}
W^\pm(U)=\frac{1}{U}\sum_{\eps\leq U}\eps g_\eps\sum_{n=1}^{\lfloor 
U/\eps\rfloor}(\pm 1)^{n-1}W^\pm(U-n\eps).
\end{equation}
 
Canonical ensemble and grand-canonical ensemble with fugacity $z=1$ 
are presently equivalent. Indeed, averaging as in \S3 the above 
relations with $q^u$ as a weight gives 
\begin{equation}\label{gencc}
Z(q)=\sum_{u\in\mathbb{U}}W(u)q^u, \quad U(q)=q\frac{d}{dq}\ln Z(q), 
\quad\sum_{\eps\in\mathbb{E}}N_\eps(q)\eps =U(q), 
\end{equation}
and  
\begin{equation}\label{Ncc}
N^+_\eps(q)=\frac{g_\eps}{q^{-\eps}-1},\qquad 
N^-_\eps(q)=\frac{g_\eps}{q^{-\eps}+1}.
\end{equation}
Inserting the latter expressions in the third equation of 
\eqref{gencc} and integrating, the expression of $\ln Z^\pm(q)$ is 
obtained from the second 
equation of \eqref{gencc} yielding  
\begin{equation}
Z^+(q)=\prod_{\eps\in\mathbb{E}}\frac{1}{(1-q^\eps)^{g_\eps}}, \qquad 
Z^-(q)=\prod_{\eps\in\mathbb{E}}(1+q^\eps)^{g_\eps}.
\end{equation}
If we expand $\ln(1\pm q^\eps)$ in power series of $q^\eps$ as was done 
earlier in \S4 we obtain 
\begin{equation}
\ln Z^\pm(q)=\sum_{n\geq 1}\frac{(\pm 1)^{n-1}}{n}Z(1,q^n). 
\end{equation}
The average energy may be derived from this expression according to 
\eqref{gencc} and the average number of particles is obtained from a 
summation of 
\eqref{Ncc} and expansion of $q^\eps/(1\pm q^\eps)$. These two 
quantities are given by  
\begin{equation}
N(q)=\sum_{n\geq 1}(\pm 1)^{n-1}Z(1,q^n),
\qquad
U(q)=\sum_{n\geq 1}(\pm 1)^{n-1}Z'(1,q^n)q^{n}. 
\end{equation}

\section{Compound systems}
A compound system is a mixture of several sub-systems comprising undistinguishable particles, the 
particles of any two sub-systems being distinguishable. We assume that 
mixing does not affect the one-particle energy-level 
values and consider two 
sub-systems only, generalization to 
a greatest number of sub-systems being straightforward.

Consider a system composed of two sub-systems with a total of
$N_1+N_2\equiv N$ particles and energy $U_1+U_2\equiv U$. If neither energy 
nor particles are allowed to be exchanged the number $W(N,U)$ of configurations of 
the compound system is the product of the statistical 
weights of the sub-systems and the occupancies add up, that is
\begin{eqnarray}
W(N,U)&=W_1(N_1,U_1)W_2(N_2,U_2),\label{mr0}\\
N_\eps(N,U)&=N_{1,\eps}(N_1,U_1)+N_{2,\eps}(N_2,U_2).
\end{eqnarray} 

Suppose now that the number of particles in each 
sub-system is constant but that some energy may be transferred from 
one sub-system to the other.  For example the two
sub-systems may comprise electrons of spin 1/2 and -1/2 respectively, spin flip 
being not allowed.  Each possible energy splitting between the two 
sub-systems has to be
taken in account.  Since all the possible configurations are
equally likely to occur any $(N_1,u_1)$-microstate of the first
sub-system has a statistical weight $W_2(N_2,U-u_1)$.  Therefore, 
the number of possible configurations of the system and the occupancy 
of level $\eps$  are 
\begin{eqnarray}\label{m1}
\fl W(N,U)&=\sum_{u_1+u_2=U}W_1(N_1,u_1)W_2(N_2,u_2), \\
\fl N_\eps(N,U)&=
\sum_{u_1+u_2=U}\frac{W_1(N_1,u_1)W_2(N_2,u_2)}{W(N,U)}
\bigl( N_{1,\eps}(N_1,u_1)+N_{2,\eps}(N_2,u_2)\bigr).\label{m2}
\end{eqnarray}
As one may expect for sub-systems that only exchange energy 
the canonical-ensemble formulas are simple. Remembering that at 
equilibrium the temperatures and chemical 
potentials of the two sub-systems equalize, the above expressions give 
 \begin{eqnarray}
Z(N,q)&=Z_1(N_1,q)Z_2(N_2,q)\label{m3},\\
N_\eps(N,q)&=N_{1,\eps}(N_1,q)+N_{2,\eps}(N_2,q)\label{m4}.
\end{eqnarray} 

If the sub-systems may exchange particles, for 
instance if spin flip is allowed in a system of electrons, 
similar arguments show that the statistical weight and the occupancy 
of level $\eps$ are 
\begin{eqnarray}
\fl W(N,U)&=\sum_{n_1+n_2=N}\sum_{u_1+u_2=U}
W_1(n_1,u_1)W_2(n_2,u_2)\label{mr1},\\
\fl N_\eps(N,U)&=\sum_{n_1+n_2=N}\sum_{u_1+u_2=U}
\frac{W_1(n_1,u_1)W_2(n_2,u_2)}{W(N,U)}
\bigl( N_{1,\eps}(n_1,u_1)+N_{2,\eps}(n_2,u_2)\bigr).\label{mr2}
\end{eqnarray} 
Partition and grand-partition functions are easily derived: 
\begin{eqnarray}
Z(N,q)&=\sum_{n_1+n_2=N}Z_1(n_1,q)Z_2(n_2,q)\label{mr3},\\
\mathcal{Z}(z,q)&=\mathcal{Z}_1(z,q)\mathcal{Z}_2(z,q).\label{mr4}
\end{eqnarray} 
Such a compound systems is equivalent to a single system 
with the following characteristics: $\mathbb{E}$ is the union of the 
sets of energy levels of the subsystems and the degeneracies are 
added. Because of the possible exchange of energy and 
particles the two sub-systems become undistinguishable from one 
another.  This fact is 
ascertained by the expression obtained for the grand-partition function of the 
compound system. But this conclusion does not apply when the sub-systems only exchange energy as illustrated in \cite{kn:acp00}.

\section{Non-degenerate system with evenly-spaced levels} 
\subsection{Microcanonical relations}
When the energy levels are evenly spaced we may assume without loss of
generality that the energy levels are non-negative integers.  We may
indeed take the energy-level spacing as the energy unit.  On the other
hand shifting the energy levels by one is tantamount to shifting the
total energy of the system by the number of particles.  The number of
configurations and the occupancies are unaffected.

We first restrict our attention to a band of evenly-spaced 
one-particle energy
levels, namely $\mathbb{E}=\{0,1,\ldots,B\}$.  Let us recall that a 
partition of
$U\in\mathbb{N}$ is a non-increasing sequence of nonnegative integers 
summing up
to $U$.  If bosons are considered, a $(N,U)$-microstate is a
partition of $U$ with at most $N$ parts none of them exceeding $B$.  
Thus the
statistical weight is the number of such partitions.  An example is 
given in
Figure~\ref{fig1}.
\begin{figure}[h]
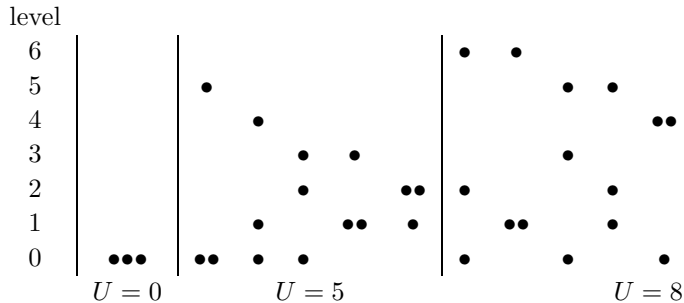
 
	\begin{center}	
	\begin{footnotesize}
	\bigskip
	\begin{tabular}{c|c|ccccc|cccccccc}
\multicolumn{1}{c}{level}&\multicolumn{14}{c}{}\\
6&&&&&&
&
$\bullet$&$\bullet$&&&&&&\\
5&&$\bullet$&&&&
&
&&$\bullet$&$\bullet$&&&&\\
4&&&$\bullet$&&&
&
&&&&$\bullet$$\bullet$&$\bullet$&$\bullet$&\\
3&&&&$\bullet$&$\bullet$&
&
&&$\bullet$&&&$\bullet$&&$\bullet$$\bullet$\\
2&&&&$\bullet$&&$\bullet$$\bullet$
&
$\bullet$&&&$\bullet$&&&$\bullet$$\bullet$&$\bullet$\\
1&&&$\bullet$&&$\bullet$$\bullet$&$\bullet$
&
&$\bullet$$\bullet$&&$\bullet$&&$\bullet$&& \\
0&$\bullet$$\bullet$$\bullet$&$\bullet$$\bullet$&$\bullet$&$\bullet$&&
&
$\bullet$&&$\bullet$&&$\bullet$&&&\\ 
\multicolumn{1}{c}{}&\multicolumn{1}{c}{$U=0$}&
\multicolumn{5}{c}{$U=5$}&\multicolumn{8}{c}{$U=8$}\\
		\end{tabular} \caption{Bosons, $(N=3,U)$-microstates for $B=6$.} 
\label{fig2}
	\end{footnotesize}
	\end{center}
\end{figure} 

\begin{figure}[h]
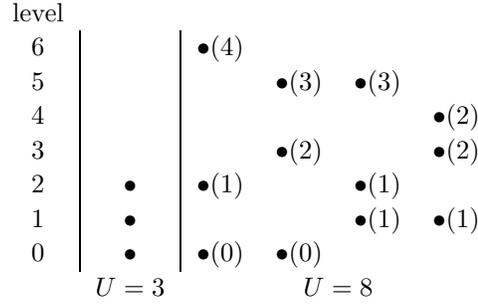
 
	\begin{center}	
	\begin{footnotesize}
	\bigskip
	\begin{tabular}{c|c|cccc}
\multicolumn{1}{c}{level}&\multicolumn{5}{c}{}\\
6&&
$\bullet$(4)&&&\\
5&&
&$\bullet$(3)&$\bullet$(3)&\\
4&&
&&&$\bullet$(2)\\
3&&
&$\bullet$(2)&&$\bullet$(2)\\
2&$\bullet$&
$\bullet$(1)&&$\bullet$(1)&\\
1&$\bullet$&
&&$\bullet$(1)&$\bullet$(1)\\
0&$\bullet$&
$\bullet$(0)&$\bullet$(0)&&\\ 
\multicolumn{1}{c}{}&\multicolumn{1}{c}{$U=3$}&
\multicolumn{4}{c}{$U=8$}\\
		\end{tabular} \caption{Fermions, $(N=3,U)$-microstates for $B=6$.} 
\label{fig3}
	\end{footnotesize}
	\end{center}
\end{figure}

\noindent With a usual notation, the statistical weight  
  \begin{equation}\label{pBNU}
 W^+(N,U)=p(B,N,U) 
\end{equation} 
vanishes when $U<0$ and $U>NB$. Several classical properties of the $p(B,N,U)$'s are given in \cite{kn:a76}. For fermions, only the partitions of $U$ with distinct parts are to be considered. Considering upward shifts from the smallest
energy $U_{\min}\equiv N(N-1)/2$ and letting $P\equiv B-N+1$ denote the 
maximum number of adjacent empty levels, it is easy to see that these partitions are 
the partitions of the \emph{added} energy $R\equiv U-U_{\min}$ into at most $N$ 
parts none of them exceeding $P$, that is 
  \begin{equation}\label{pPNR}
 W^-(N,U)=p(B-N+1,N,U-N(N-1)/2)=p(P,N,R). 
\end{equation}
 Figure~\ref{fig3} illustrates the above argument with reference to
 Figure~\ref{fig2}.  The numbers in parentheses count the number of energy 
 jumps from the initial positions and relate the fermionic (3,8)-microstates to the 
corresponding bosonic (3,5)-microstates.

In the present situation of evenly-spaced levels further micro-canonical 
relations may be established.  The
following reasoning may be formalized by using the definition in 
\eqref{defm} of
microstates.  Consider first a bosonic system consisting of $N$ particles and 
energy $U$. 
Recall that $W^+(N-1,U-\eps)$ is the number of $(N,U)$-microstates 
with at least
one particle at level $\eps$. There are therefore $W^+(N,U)-W^+(N-1,U)$ 
microstates
with no particle at level 0.  Shifting downward the energy levels by 1 these microstates  
provide all the $(N,U-N)$-microstates except those that have one
particle at least at level $B$.  In other words
\begin{equation}\label{recW+}
W^+(N,U)-W^+(N-1,U)=W^+(N,U-N)-W^+(N-1,U-N-B).
\end{equation}
Let us denote for short $M_k(N,U)\equiv W(N,U)N_k(N,U)$.  
Substituting $N-n$ and
$U-n(k+1)$ in place of $N$ and $U$ in \eqref{recW+} and summing both sides 
with $n$
running from 0 to $N$, we obtain from \eqref{mco} a relation 
involving the same
number $N$ of particles but successive levels $k,k+1$.  This relation reads for $0\leq 
k<B$ 
\begin{equation}\label{reckN+}
\begin{array}{ll}
W^+(N,U)+M_{k+1}^+(N,U)-M_{k+1}^+(N,U+k+1)\quad =  \\
W^+(N,U-N)+M_k^+(N,U-N)-M_k^+(N,U-N-B+k). 
\end{array}
\end{equation}
Indeed, using \eqref{mco} we have successively
\begin{eqnarray*}
\fl\sum_{n\geq 0}W^+(N-n,U-n(k+1))&=W^+(N,U)+W^+(N,U)N_{k+1}^+(N,U),\\
\fl\sum_{n\geq 0}W^+(N-1-n,U-n(k+1))&=\sum_{n\geq 1}W^+(N-n,U+k+1-n(k+1))\\
&=W^+(N,U+k+1)N_{k+1}^+(N,U+k+1),
\end{eqnarray*}
the right-hand terms in \eqref{recW+} being handled analogously. 

Consider now a fermionic system consisting of $N$ particles and with 
total energy $U$.  We 
have 
\begin{equation}\label{recW-}
\fl W^-(N,U)+W^-(N-1,U-B-1)=W^-(N,U-N)+W^-(N-1,U-N+1). 
\end{equation}
Indeed, the left-hand side of this expression counts the union of the 
$(N,U)$-microstates and the
fictitious $(N,U)$-configurations with 1 particle at the fictitious 
additional level
$B+1$.  Depending on whether the above $(N,U)$-configurations have 0 
particle at
level 0 or not, shifting downward the levels by 1 results either in a
$(N,U-N)$-microstate or in a $(N,U-N)$-configuration with 1 particle at 
the
fictitious
level $-1$.  This gives the right-hand side. Notice further that 
analogously
the occupancy of the lowest level satisfies the relation
\begin{equation}\label{recinit}
M_0(N,U)+M_0(N-1,U-B-1)=W^-(N-1,U-N+1),
\end{equation}
because shifting the levels yields all the $(N,U)$-configurations 
with 1
particle at level $-1$.  In the same way as in \eqref{reckN+} one derives 
from
\eqref{recW+}, we obtain from \eqref{recW-} for $0\leq k<B$  
\begin{equation}\label{reckN-}
\begin{array}{ll}
W^-(N,U)-M_{k+1}^-(N,U)+M_{k+1}^-(N,U-B+k)\quad =  \\
W^-(N,U-N)-M_k^-(N,U-N)+M_k^-(N,U-N+1+k). 
\end{array}
\end{equation}
Replacing $M_0(N,U)$ in \eqref{recinit} by its value given by
\eqref{mcor} and substituting $N$ to $N-1$ we obtain
\[
M_0(N,U)=M_0(N,U-B-1)+W^-(N,U)-W^-(N,U-N). 
\]
This relation shows that formula \eqref{reckN-} still holds for $k+1=0$. 

\subsection{Canonical relations}
It is well known that the generating series for $U$ of the 
numbers $p(B,N,U)$ in \eqref{pBNU} is the
gaussian polynomial
\[
\sum_{U=0}^{NB}p(B,N,U)q^U
=\frac{(q)_{B+N}}{(q)_{B}(q)_{N}}, 
\]
where $(q)_{n}$ is the commonly employed notation for the product 
$(1-q)(1-q^2)\ldots(1-q^n)$ and $(q)_{0}=1$.  Accordingly,
\begin{equation}\label{hZ}
Z^+(N,q)=\frac{(q)_{B+N}}{(q)_B(q)_N},\qquad 
Z^-(N,q)=q^{ N(N-1)/2}\frac{(q)_{B+1}}{(q)_{B+1-N}(q)_N}. 
\end{equation}
In order to keep the paper self-contained let us point out that the 
latter expressions may be easily derived from the previous microcanonical relations.  
According to \eqref{pBNU} and \eqref{pPNR} it suffices to consider $Z^+(N,q)$.  Firstly
\[
Z(1,q)=1+q+\cdots +q^B=\frac{1-q^{B+1}}{1-q}.
\]  
Secondly, averaging \eqref{recW+} with $q^U$ as a weight and 
collecting the results we obtain the recurrence relation
\[
Z^+(N,q)=\frac{1-q^{B+N}}{1-q^N}Z^+(N-1,q), 
\]
from which the derivation of  $Z^+(N,q)$ in \eqref{hZ} is 
straightforward. 

Averaging \eqref{reckN+} and \eqref{reckN-} with $q^U$ as a 
weight, dividing by $Z(N,q)$ and multiplying by $q^{k+1}$, 
provides simple recursive formulas for the computation 
of canonical occupancies. These formulas are new for bosons, and generalize \cite{kn:s00} for fermions. For $0\leq k<B$ they are 
\begin{eqnarray}
\fl (1-q^{k+1})N^+_{k+1}(N,q)&=
q^{k+1}(1-q^N)-q^N(q^{k+1}-q^{B+1})N^+_k(N,q),\label{reckN+c}\\
\fl (q^{k+1}-q^{B+1})N^-_{k+1}(N,q)&=
q^{k+1}(1-q^N)-q^N(1-q^{k+1})N^-_k(N,q)\label{reckN-c}.
\end{eqnarray}
As noticed after \eqref{reckN-} the latter formula holds for $k+1=0$ so that 
\[
N^-_0(N,q)=\frac{1-q^N}{1-q^{B+1}}. 
\]
There is apparently no such simple expression for $N^+_0(N,q)$. This 
quantity may be calculated from \eqref{cor} with the initial value 
$N^+_0(1,q)=1/Z(1,q)$. The expression obtained directly from \eqref{co} is 
\[
N^+_0(1,q)=\frac{(q)_N}{(q)_{N+B}}\sum_{n=0}^{N-1}\frac{(q)_{B+n}}{(q)_n}.
\]

\subsection{Unbounded energy levels.}
If we let $B\rightarrow\infty$ in the previous formulas we obtain 
expressions applicable to infinitely many energy levels 
\begin{eqnarray*}
W^+(N,U)=p_N(U),   \quad &W^-(N,U)=p_N(U-N(N-1)/2)=p_N(R),\\
Z^+(N,q)=\frac{1}{(q)_N}, &
Z^-(N,q)=\frac{q^{ N(N-1)/2}}{(q)_N}. 
\end{eqnarray*}
The recursive formulas obtained in the last 
subsection were given before in \cite{kn:s00} for fermionic systems in the limit of infinite $B$-values.

If we introduce the additional energy $R$ in place of the total 
energy $U$ in fermionic systems the statistical weights and 
partition 
functions exhibit the same form for bosons and fermions.  This 
explains the similarities previously noted, for instance 
the fact that Bose and Fermi gases (with the same number $N$ of 
particles) in square-law potentials have the same heat capacity.  
Considering $R$ instead of $U$ enables one to let 
$N\rightarrow\infty$ in 
the case of fermionic systems \cite{kn:sm96,kn:abcp99,kn:p00,kn:bp00}. This limit 
provides with the simplest form of all previous formulas. 

\section{Conclusion}
Non-interacting fermions and bosons statistics have been obtained in the three classical ensembles. The originality of the method rests on the derivation of complete and explicit microcanonical formulas by direct enumeration of the microstates. Canonical and grand-canonical relations are further derived by simple averaging. Several results are new, older works are referred to. 

Using the microcanonical relations requires the sole knowledge of system energy levels and degeneracies. Only average occupation numbers have been reported, but formulas for joint and higher moments may be obtained by similar arguments \cite{kn:k51}. Microcanonical formulas have been given that arise naturally. One of them leads to a  recursive relation for the derivative of the canonical partition function. 
We have also shown that most canonical and grand-canonical quantities may be calculated from the one-particle canonical partition functions at temperatures $T/n$, $n\geq 1$. This result is still valid for particles that do not bear generalized charges. 

One-dimensional harmonic oscillators Schönhammer's results \cite{kn:s00} have been generalized in several aspects. Bosons are also treated. Related microcanonical relations are given. The case of finitely many levels is treated. Extension to arbitrary degeneracy shall be the object of future research. 

Other, less classical, statistical ensembles might be handled analogously. For example, in the grand microcanonical ensemble \cite{kn:lk81} the system may exchange particles with a bath. The fugacity $z$ and the energy $U$ of the system are fixed but the number $n$ of its particles is allowed to fluctuate. The probability that a particular $(n,U)$-microstate occurs is proportional to $z^n$. Accordingly, the grand microcanonical statistics can be derived from averaging the microcanonical relations in \S 2 with $z^n$ as a weight. 

\appendix
\section{Relations between fermionic and bosonic statistical 
weights.}
Given the set $\mathbb{E}$ of the one-particle energy levels, there is
a simple relation between the numbers $W^- (N,U)$ and $W^+ (N,U)$.  
It reads
\begin{equation}\label{Wfb}
W^- (N,U)=\sum_{n=0}^{\lfloor N/2\rfloor}(-1)^n
\sum_{u=0}^{U/2}W^-(n,u)W^+ (N-2n,U-2u),  	
\end{equation}
where $\lfloor N/2\rfloor$ denotes the greatest integer non-exceeding 
$N/2$, and
the second $\sum$ symbol stands for a summation for all values of
$u\in\mathbb{U}$ between $0$ and $U/2$.  Let us sketch a proof of 
relation
\eqref{Wfb}.  Suppose the Pauli exclusion principle applies to all energy 
levels except
$\eps$.  Then the possible $(N,U)$-configurations split up into those 
that obey
strictly  the Pauli principle and those that exhibit at least two 
particles
at level $\eps$.  Removing two particles at level $\eps$ in the 
latter, denoting
by $W^{\eps}(N,U)$ the number of $(N,U)$-configurations of our 
supposed system
and enumerating, result in
\begin{equation}\label{prov1} 
W^-(N,U)=W^{\eps}(N,U)-W^{\eps}(N-2,U-2\eps).
\end{equation}
Further relaxing the Pauli principle at other levels one by one and 
iterating 
formula \eqref{prov1}, we get 
\begin{equation}\label{prov2}
W^-(N,U)=\sum_{E\subseteq\mathbb{E}}(-1)^{|E|}
W^+(N-2|E|,U-2\textstyle{\sum_{\eps\in E}}\eps), 
\end{equation}
where $|E|$ stands for the number of levels in $E$.  Finally 
formula \eqref{prov2} may alternatively be written as formula \eqref{Wfb} by 
letting $n=|E|$, $u=\sum_{\eps\in E}\eps$, and remarking that there 
are $W^-(n,u)$ ways to get $u$ from the $n$ elements in $E$.  All 
sums terminate because $W(N,U)=0$ for negative $N$ or $U$. 

Averaging formula \eqref{Wfb} we readily derive the identity 
\begin{equation}\label{Zfb}
Z^-(N,q)=\sum_{n=0}^{\lfloor N/2\rfloor}(-1)^nZ^+(N-2n,q)Z^-(n,q^2). 	
\end{equation}
Denoting by $\mathcal{Z}^\pm (z,q)$ the grand-canonical partition 
function further averaging yields 
\[
\mathcal{Z}^-(z,q)=\mathcal{Z}^+(z,q)\mathcal{Z}^-(-z^2,q^2),
\]
a fairly straightforward identity in view of \eqref{sgqtW}. 
In the present case we recognize that it is easier to proceed the other 
way around, namely from
grand-canonical to microcanonical relations.  As an illustration, 
from the similar identities 
\[
1=\mathcal{Z}^+(z,q)\mathcal{Z}^-(-z,q),\quad\quad 
\mathcal{Z}^+(z,q)=\mathcal{Z}^-(z,q)\mathcal{Z}^+(z^2,q^2),
\]
we derive by expanding the series products and identifying the coefficients of $z^N$, 
for positive $N$, 
\begin{eqnarray*}
0&=\sum_{n=0}^N(-1)^nZ^+(n,q)Z^-(N-n,q),\\ 
Z^+(N,q)&=\sum_{n=0}^{\lfloor N/2\rfloor}Z^-(N-2n,q)Z^+(n,q^2).
\end{eqnarray*}
The microcanonical counterparts of these relations are obtained by identifying the 
coefficients of $q^U$. 

\section{Entropy}
For the reader's convenience we link in what follows statistical quantities with 
thermodynamical ones. This can be done by examinating the expression of entropy in both formalisms. The von Neumann-Shannon entropy is defined as 
\[
S=-\sum_{\omega}p_\omega\ln p_\omega,
\]
where $p_\omega$ is the probability of a state of the system and 
$\omega$ runs through the set of all such states. 

In the micro-canonical ensemble with $N$ particles and total energy 
$U$, the microstates are equally likely so that 
$p_{\omega(N,U)}=1/W(N,U)$ and $S=\ln W(N,U)$, namely the Boltzmann 
entropy. 

In the canonical ensemble with $N$ particles and temperature $T$, 
$p_{\omega(N,u)}=q^u/Z(N,q)$. Thus 
\begin{eqnarray*}
S&=-\sum_{u\in\mathbb{U}}W(N,u)\frac{q^u}{Z(N,q)}\ln\frac{q^u}{Z(N,q)}
=\ln Z(N,q)-\frac{q\ln q}{Z(N,q)}\frac{d}{dq}Z(N,q)\\
&=\ln Z(N,q)+\frac{1}{k_{\mathrm{B}}T}U(N,q).
\end{eqnarray*}
This is the classical expression of the entropy of a closed 
thermodynamic system with Helmholz free energy $A(N,q)=-k_{\mathrm{B}}T\ln 
Z(N,q)$. 

In the grand-canonical ensemble with fugacity $z$ and temperature 
$T$, the probability $p_{\omega(n,u)}=z^nq^u/\mathcal{Z}(z,q)$. Accordingly, we obtain as above 
\[\fl 
S=-\sum_{u\in\mathbb{U}}\sum_{n\geq 
0}W(n,u)\frac{z^nq^u}{\mathcal{Z}(z,q)}\ln\frac{z^nq^u}{\mathcal{Z}(z,q)}=
\ln\mathcal{Z}(z,q)+\frac{1}{k_{\mathrm{B}}T}U(z,q)-\frac{\mu}{k_{\mathrm{B}}T}N(z,q).
\]
The relevant thermodynamical potential (grand potential) is given by the logarithm of the partition function. Accordingly, entropies of subsystems add up if their partition functions multiply. The appropriate conditions are given in Section 6.

\end{document}